\documentclass[aps,prb,groupedaddress,twocolumn,showpacs]{revtex4}
\usepackage{graphicx}

\begin{document}
\title{Resistivity scaling and critical dynamics of fully frustrated
Josephson-junction arrays with on-site dissipation }
\author{Enzo Granato}
\address{Laborat\'orio Associado de Sensores e Materiais,
Instituto Nacional de Pesquisas Espaciais, 12201-190 S\~{a}o
Jos\'e dos Campos, S\~ao Paulo, Brazil}
\author{Daniel Dom\'{\i}nguez}
\address{Centro At\'omico Bariloche and Instituto Balseiro, 8400 San Carlos de Bariloche,
Rio Negro, Argentina}

\begin{abstract}
We study the scaling behavior and critical dynamics of the
resistive transition in Josephson-junction arrays, at $f=1/2$ flux
quantum per plaquette, by numerical simulation of an on-site
dissipation model for the dynamics. The results are compared with
recent simulations using the resistively-shunted-junction model.
For both models, we find that the resistivity scaling and critical
dynamics of the phases are well described by the same critical
temperature as for the chiral (vortex-lattice) transition, with a
power-law divergent correlation length. The behavior is consistent
with the single transition scenario, where phase and chiral
variables order at the same temperature, but with different
dynamic exponents $z$ for phase coherence and chiral order.

\end{abstract}


\maketitle

\section{Introduction}

There has been considerable interest, both experimentally and
theoretically, in phase transitions of two-dimensional Josephson
junction arrays (JJA). \cite{conf,zant,carini,martinoli,theron,%
yun,yu,ling,beck,tiesinga,kim99,%
jensen,shenoy,minhagen,wallin,teitel,rg,gkn,lkg,rj,gkn96,knops,leelee,olsson,%
nicodemi,honda,diep,korshu,eta,modjja,%
zheng,ozeki,teitel89,simkin98,leeteitel,gks98,md01,gd03,dd99,kim99b,eg98}
Such arrays can be artificially fabricated as a lattice of
superconducting grains connected by an insulator or a normal metal
\cite{conf,zant,carini,martinoli,theron,yun} and are also closely
related to superconducting wire networks.\cite{yu,ling}
Experimentally, the resistive transition has been the one most
extensively studied, \cite{zant,carini,martinoli,yun,yu,ling}
while theoretically several studies of  XY models,
\cite{beck,tiesinga,kim99,jensen,%
shenoy,minhagen,wallin,teitel,rg,gkn,lkg,rj,gkn96,knops,leelee,olsson,%
nicodemi,honda,diep,korshu,eta,modjja,%
zheng,ozeki,teitel89,simkin98,leeteitel,gks98,md01,gd03,dd99,kim99b,eg98}
which describe the ideal JJA, have been done. A significant
understanding of these systems has already been  achieved by
comparing the results of experiments with the theoretically
predicted equilibrium critical behavior, with and without an
applied magnetic field. However, to a large extent, dynamical
critical behavior remains much less understood, particularly in
the presence of a magnetic field, where frustration effects may
introduce additional elementary excitations relevant for the
static and dynamic critical behavior. It is well known that while
static critical phenomena depend on the spatial dimensionality as
well as on the symmetry of the order parameter, the dynamic
universality class of the phase transition will depend upon
additional properties which do not affect the statics as, for
example, conservation laws for the order parameter. \cite{hh}
Thus, testing the universality hypothesis of dynamical critical
behavior requires the study of specific dynamical models. For JJA,
the physically relevant dynamical model for the phase dynamics can
not be unambiguously identified
\cite{theron,beck,tiesinga,kim99,jensen} and  should depend on the
particular coupling mechanism between the superconducting elements
of the array. It is expected that, at least for an array of ideal
tunnel junctions, the Resistively-Shunted-Junction (RSJ) model of
current flow between superconducting grains would be a more
physical representation of the system.\cite{shenoy} This model
assumes that energy dissipation occurs only through the junctions
and imposes current conservation at each site of the array. On the
other hand, for wire networks or arrays of proximity-effect
junctions, local dissipation at the sites of the array should also
be allowed leading to a  model with  on-site dissipation for the
dynamics.

In experimental investigations of JJA,, the resistive transition
is usually identified from the behavior of the current-voltage
(IV) characteristics near the critical temperature. The divergent
correlation length determines both the linear and nonlinear
resistivity sufficiently close to the transition. To interpret the
data and determine the underlying equilibrium transition, the
scaling theory of Fisher {\it et al.} \cite{ffh} has been widely
used. For JJA at zero magnetic field, which is isomorphic to the
standard XY model, the resistive transition is in the
Kosterlitz-Thouless (KT) universality class,
\cite{conf,shenoy,minhagen} where the correlation length diverges
exponentially near the critical temperature.  Studies of the
critical dynamical behavior using Monte Carlo (MC) dynamics
\cite{wallin} and RSJ or on-site dissipation dynamics,
\cite{kim99,jensen} find a behavior consistent with the dynamical
theory of the KT transition. The exponent of the current-voltage
relation, $V\sim I^a$, at the transition, assuming the universal
value $a=z+1=3$, corresponds to a dynamic exponent $z=2$ in  the
resistivity scaling theory. \cite{ffh}

However, in frustrated  Josephson-junction arrays (FJJA),
corresponding to $f=1/2$ flux quantum per plaquette, besides the
phase variables, the vortex-lattice induced by the external field
introduces an additional discrete (Ising-like) order parameter,
the chirality,\cite{teitel} which measures the direction of local
current circulation in the array. The ground state consists of a
pinned commensurate vortex-lattice corresponding to an
antiferromagnetic arrangement of chiralities and vortex-lattice
melting corresponds to the chiral order-disorder transition. As a
consequence, two distinct scenarios for the occurrence of phase
transitions as a function of temperature have been proposed by
Teitel and Jayaprakash (Ref. \onlinecite{teitel}(b)): separated
chiral and phase-coherence transitions or a single transition
where phase and chirality order at the same temperature. In the
former scenario, the phase transitions should be in the KT and
Ising universality classes, respectively, while in the later
scenario the critical behavior should be a superposition of KT and
Ising critical behavior at the same critical temperature, if the
coupling between phase and chiral variables are irrelevant at
criticality (decoupled single transition), otherwise new critical
behavior (coupled single transition) should occur with phase
coherence and chiral order showing critical behavior different
from the KT and Ising universality classes. These possible
scenarios are supported, for example, by renormalization-group
studies based on the Ginzburg-Landau expansion of the frustrated
XY model (FXY) \cite{rg} which also shows that the universality
class of these transitions can be described by the XY-Ising model
\cite{gkn}. It appears that the current predominant point of view
is that the separated transition scenario is realized with a KT
transition  occurring below the chiral transition. Recently, this
scenario has received further support from  appealing arguments by
Korshunov \cite{korshu}, based on chiral domain wall fluctuations
and vortex unbinding, which provides a mechanism for the
separation of the two transitions in this order.
Also, there are significant numerical evidences from equilibrium
calculations which favor this scenario.
However, the coupled single transition scenario has also received
some support from different calculations of the chiral critical
exponents and central charge from finite-size scaling which show
results different from the pure Ising values, but several of these
studies do not verify if the transition temperature for
phase-coherence coincides with the chiral transition temperature.
On the other hand, the numerical evidence for the separated
transitions scenario finds that the phase-coherence transition
\cite{olsson} is consistent with KT behavior but the critical
exponents found for the chiral transition by finite-size scaling
do not show the expected pure Ising behavior according to Ref.
\onlinecite{leelee}. However, it has  been found by Olsson
\cite{olsson} that the value of the thermal critical exponent is
consistent with pure Ising value depending on the temperature
region in which a fit is made. Therefore, the deviations of the
exponents from pure Ising values can not be regarded as an
unambiguous evidence for non-Ising critical behavior. The
separated transition scenario also relies on the assumption that
the phase-coherence transition is pure KT and therefore uses some
of the predicted behavior from the KT theory, like the helicity
modulus jump or exponentially divergent correlation length, to
locate this critical temperature. If the helicity modulus jump is
actually larger than the universal value then the procedure of
locating the critical temperature from the jump \cite{olsson} can
only overestimate the critical temperature. Although this
assumption is consistent with a phase-coherence critical
temperature  below the chiral transition, such procedure could
result in an underestimate of the phase-coherence critical
temperature if the transitions coincide or the chiral transition
occurs below the phase-coherence transition. In fact, it has  been
shown that if one enlarges the parameter space of the FXY model
\cite{eta} by considering a model where every other column in the
square lattice has coupling constants which differ from the others
by a constant ratio $\rho$, the chiral transition occurs below the
phase-coherence transition \cite{modjja} if $\rho$ is sufficiently
different from $1$. It is then found that there is a singular
contribution to the temperature dependence of the helicity modulus
near the chiral transition \cite{eta} determined by the chiral
critical exponents. For the standard FXY model, obtained when
$\rho \rightarrow 1$, such singular contribution will remain if
the transition is single and therefore it can affect the helicity
modulus behavior near the transition.

In any case, independently of the scenario interpretation, several
numerical calculations using quite different methods agree
\cite{rj,leelee,olsson,zheng,ozeki} with the earlier estimate of
the chiral transition temperature \cite{lkg} at $T_{ch}=0.455$
within a $0.8\% $ errorbar. On the other hand, for the
phase-coherence transition, it is clear that it would be more
satisfactorily if it could be determined by methods which do no
rely on assumptions of KT behavior.

These different phase-transition scenarios have important
consequences for the resistive behavior of the FJJA. Since the
resistive transition corresponds to the onset of phase coherence,
they imply quite distinct behavior. In the separated transition
scenario or single but decoupled scenario, the resistive behavior
should be described by the KT universality class. On the other
hand, in the single coupled scenario, where the critical dynamics
involve strongly coupled phase and chiral variables, the resistive
behavior should be significantly different. In principle, such
behavior can be detected experimentally.

Measurements of current-voltage curves in FJJA were  fitted
assuming pure KT behavior,\cite{zant,carini} but either an
unexpectedly low value of the transition temperature (compared
with theoretical expectations) was obtained in one case
\cite{zant} or the IV exponent at the transition was $a < 3$ in
the other case \cite{carini}. More recently, the current-voltage
curves in JJA\cite{yun} and in superconducting
networks\cite{yu,ling} were found to be better described by a
power-law correlation length. However, very different values of
the critical exponents $z,\nu$ were obtained in each case.

Earlier numerical studies of the IV characteristics for FJJA,
obtained with RSJ dynamics \cite{teitel89,simkin98} or  MC
dynamics, \cite{leeteitel} were performed  for small system sizes
($L\le 16$). In particular, the studies with RSJ dynamics used
free boundary conditions to impose a driving current. This leads
to significant additional dissipation due to boundary effects,
\cite{gks98} specially in small system sizes.  Other works have
studied the short-time dynamics of chirality, \cite{zheng} and the
non-equilibrium transitions at large currents.\cite{md01}

Recently,\cite{gd03} we have studied  the critical dynamics and
resistivity scaling in FJJA by numerical simulation of the RSJ
dynamics with periodic (fluctuating twist) boundary conditions
including much large systems sizes. It was found that the
current-voltage scaling is consistent with the single-transition
scenario. The scaling behavior is well described by a resistive
transition occurring at a critical temperature corresponding to
the chiral transition, with a power-law divergent correlation
length, but with two different dynamic exponents, $z_{ph}\sim 1$
and $z_{ch}\sim 2$, for phase and chiral variables, respectively.
This result implies that,  at the transition, the exponent of the
IV power-law, $V\sim I^a$, is $a = z_{ph}+1 \approx 2$ rather than
$a=3$ as for the unfrustrated case. In view of the possible
dependence of the dynamic behavior on the particular RSJ dynamics
used in these simulations, it should also be of interest to study
the resistive behavior with an on-site dissipation model for the
dynamics. Results for this dynamical model should be particularly
relevant for frustrated wire networks \cite{ling} or
proximity-effect junctions.

In this work we study the resistivity scaling and critical
dynamics of a frustrated Josephson-junction array, defined on
square lattice, at $f=1/2$ flux quantum per plaquette, by
numerical simulations of an on-site dissipation model for the
array dynamics. Using a dynamic scaling analysis, we find that the
resistivity behavior and critical dynamics are well described by
the critical temperature corresponding to the chiral
(vortex-lattice) transition with a correlation length that
diverges as a power law. Two dynamic exponents, $z_{ph}\sim 1.5$
and $z_{ch}\sim 2.5$, are found for phase-coherence and chiral
order, respectively. Consequently, at the transition, the exponent
of the current-voltage power-law, $V\sim I^a$, is $a = z_{ph}+1
\approx 2.5$ rather than $a=3$ as for the unfrustrated case. This
is the same behavior we have found recently for the RSJ model
\cite{gd03} but with different values for the dynamic exponents
($z_{ph}\sim 0.9(1)$ and $z_{ch}\sim 2.1$). Including on-site
dissipation in the dynamical model could be a more realistic
description of wire networks than the RSJ model. Indeed,
resistivity scaling of experimental data on wire networks
\cite{ling} find $z \sim 2$, which is consistent with our estimate
$z_{ph}$ within the experimental errors, and also shows that the
resistivity scaling is well described by a power-law correlation
length as found in our simulations.

\section{Model and simulation}

The hamiltonian of  a square two-dimensional array under  a
magnetic field is given by
\begin{equation}
\mathcal{H} = -E_J\sum_{\mathbf{r},\mathbf{\mu}}
\cos(\theta_{\mathbf{r}+\mathbf{\mu}}-\theta_\mathbf{r}-A_{\mathbf{r},\mathbf{\mu}})
\label{hamilt1}
\end{equation}
where $\theta_\mathbf{r}$ is the superconducting phase of the
grain at site $\mathbf{r} = (n_x a, n_y a)$  with  $n_x,n_y$
integers, and $a$ the lattice constant, and $\mathbf{\mu} =
\mathbf{x},\mathbf{y}$ with $\mathbf{x}=(a,0)$,
$\mathbf{y}=(0,a)$, and $E_J=I_0\hbar/2e$ the Josephson energy.
The magnetic field introduces frustration through the vector
potential integral
\begin{equation}
A_{\mathbf{r},\mathbf{\mu}}=
\frac{2\pi}{\Phi_0}\int_\mathbf{r}^{\mathbf{r}+\mathbf{\mu}}
\mathbf{A}\cdot d\mathbf{l}\;,
\end{equation}
which satisfies
\begin{eqnarray}
\Delta_{\mu}\times A_\mathbf{r,\mu}&=&A_{\mathbf{r},x}-A_{{\bf
n}+{\bf y},x}+
A_{{\bf r}+{\bf x},y}-A_{{\bf r},y}\nonumber\\
 &=&2\pi f,
\end{eqnarray}
with $f=H a^2/\Phi_0$, where $H$ is the applied magnetic field and
$\Phi_0=h/2e$ the quantum of flux. The fully frustrated case
corresponds to half quantum of flux per plaquette, $f=1/2$.

The simulations are performed with the same ``fluctuating twist"
boundary conditions as used, for example, in Refs.
\onlinecite{kim99,dd99,kim99b}. This consists on considering
periodic boundary conditions for  the supercurrents in the
$\mathbf{\mu}$ direction
 while adding a fluctuating twist
$\alpha_\mu$ to the gauge invariant phase in the
$\mathbf{\mu}$ direction.
In this case the gauge invariant phase difference is modified to
\begin{equation}
\theta_{\mathbf{r},\mathbf{\mu}}=
\theta_{\mathbf{r}+\mathbf{\mu}}-\theta_\mathbf{r}-A_{\mathbf{r},\mathbf{\mu}}+\alpha_\mu
\end{equation}
For the vector potential we choose the Landau gauge
\begin{eqnarray}
A_{\mathbf{r},x}&=&-2\pi f n_y\nonumber\\
A_{\mathbf{r},y}&=&0
\end{eqnarray}
In this gauge, the boundary condition for the phases in a system
of size $L \times L$ is given by
\begin{eqnarray}
\theta(n_x+L,n_y)&=&\theta(n_x,n_y)  \nonumber \\
\theta(n_x,n_y+L)&=&\theta(n_x,n_y)-2\pi f Ln_x\; .
\end{eqnarray}
For $f=1/2$ and $L$ even, the second condition is irrelevant, but
not for general frustration $f$. In the presence of an external
current $I_\mathrm{ext}^\mu$ in the $\mathbf{\mu}$ direction, one
has to add the term $-\frac{\hbar}{2e}L^2 I_\mathrm{ext}^\mu
\alpha_\mu$ in the hamiltonian of Eq. (\ref{hamilt1}), which
couples the current with the global phase difference per row,
$L\alpha_\mu$, introduced by the fluctuating twist. Therefore, the
hamiltonian of a frustrated square array with fluctuating twist
boundary conditions and an external current is
\begin{eqnarray}
\mathcal{H} &=& -E_J\sum_{\mathbf{r},\mathbf{\mu}}
\cos(\theta_{\mathbf{r}+\mathbf{\mu}}-\theta_\mathbf{r}-A_{\mathbf{r},\mathbf{\mu}}+\alpha_\mu)
\nonumber\\
& &- \frac{\hbar}{2e}L^2\sum_{\mathbf{\mu}}I_\mathrm{ext}^\mu
\alpha_\mu \label{hamilt2}
\end{eqnarray}

We define the on-site dissipation dynamics by considering the
local  Langevin equations for the  fluctuating variables
$\theta_\mathbf{r}$ and $\alpha_\mathbf{\mu}$
\begin{eqnarray}
\frac{d\theta_\mathbf{r}}{dt} &= &-\Gamma_\theta
\frac{\delta\mathcal{H}}{\delta\theta_\mathbf{r}}
+\eta_\mathbf{r}(t)\label{dtheta}\\
\frac{d\alpha_\mathbf{\mu}}{dt} &= &-\Gamma_\alpha
\frac{\delta\mathcal{H}}{\delta\alpha_\mathbf{\mu}} +\eta_\mu(t)
\label{dalpha}
\end{eqnarray}
where $\Gamma_\theta$, $\Gamma_\alpha$ are dissipation parameters,
and the noise terms have zero average and correlations
\begin{eqnarray}
\left\langle
\eta_\mathbf{r}(t)\eta_\mathbf{r'}(t')\right\rangle&=&
2k_BT\, \Gamma_\theta\, \delta_{\mathbf{r},\mathbf{r'}}\delta(t-t')\\
\left\langle
\eta_\mathbf{\mu}(t)\eta_\mathbf{\mu'}(t')\right\rangle&=& 2k_BT\,
\Gamma_\alpha\, \delta_{\mathbf{\mu},\mathbf{\mu'}}\delta(t-t')
\end{eqnarray}
The dissipation constant $\Gamma_\alpha$ should be proportional to
$L^{-2}$ in order to be an intensive quantity. A convenient choice
is
$$\Gamma_\alpha=\frac{\Gamma_\theta}{L^2}$$
(in general it can be $\Gamma_\alpha=\beta\Gamma_\theta/L^2$, here
we choose $\beta=1$ to be consistent with Ref.
\onlinecite{jensen}).

Dimensionless quantities are used with time in units of $\tau =
2e/\hbar\Gamma_\theta I_{0}$, currents in units of $I_{0}$,
voltages in units of $(\hbar/2e)^2\Gamma_\theta I_{o} $ and
temperature in units of $\hbar I_{o}/2ek_{B}$. A total current $I$
is imposed uniformly in the array in the $\mathbf{y}$-direction
 with current density $J=I/L$,
where $L$ is the system size and the average electric field $E$ is
obtained from the voltage $V$ across the system as
$E=V/L=(\hbar/2e)\langle{d\alpha_y/dt}\rangle$, where $\alpha_y L$
is the global phase difference or twist in the
$\mathbf{y}$-direction. With all this considerations, the
dimensionless equations of motion are then
\begin{eqnarray}
\frac{d\theta_\mathbf{r}}{dt} &= & -\Delta_{\mu}\cdot
S_{\mathbf{r},\mathbf{\mu}}
+\eta_\mathbf{r}(t)\\
\frac{d\alpha_\mathbf{\mu}}{dt} &=
&-\frac{1}{L^2}\sum_{\mathbf{r}} S_{\mathbf{r},\mathbf{\mu}} + I
\delta_{\mu,y} + \eta_\mu(t)
\end{eqnarray}
where the supercurrent  is defined by
$$S_{\mathbf{r},\mathbf{\mu}}=\sin(\theta_{\mathbf{r}+\mathbf{\mu}}-\theta_\mathbf{r}
-A_{\mathbf{r},\mathbf{\mu}}+\alpha_\mu)\;,$$ and the discrete
divergence operator is defined as
$$\Delta_{\mu}\cdot S_{\mathbf{r},\mathbf{\mu}}=
\sum_{\mu=x,y} S_{\mathbf{r},\mathbf{\mu}} -
S_{\mathbf{r}-\mathbf{\mu},\mathbf{\mu}} . $$ Finally, the now
dimensionless noise variables $ \eta_\mathbf{r}(t)$ have
correlations
\begin{eqnarray}
\left\langle
\eta_\mathbf{r}(t)\eta_\mathbf{r'}(t')\right\rangle&=&
2T\, \delta_{\mathbf{r},\mathbf{r'}}\delta(t-t)\\
\left\langle
\eta_\mathbf{\mu}(t)\eta_\mathbf{\mu'}(t')\right\rangle&=&
\frac{2T}{L^2}\, \delta_{\mathbf{\mu},\mathbf{\mu'}}\delta(t-t)\,.
\end{eqnarray}

The set of Eqs (\ref{dtheta})-(\ref{dalpha}) describe the dynamics
of JJA with "on-site dissipation"  in contrast to the RSJ dynamics
which only considers dissipation through the junctions
\cite{shenoy}. The on-site dissipation dynamical model has been
studied previously \cite{beck,kim99,jensen,kim99b} for the
unfrustrated ($f=0$) case and compared with the RSJ dynamics.
Their main difference is that while the on-site dynamics
corresponds to a local damping the RSJ dynamics corresponds to a
nonlocal damping \cite{beck,shenoy}. A physical interpretation of
the on-site dynamics for JJA in terms of currents and voltages has
also been discussed previously \cite{beck,kim99,jensen,kim99b}.
Its main features are summarized in the following.   (i) It takes
into account normal current flow between each superconducting node
and the substrate, which leads to a current leakage through a
resistance  to the ground $R_0$. (ii) It neglects the
quasiparticle normal current of each junction, which is associated
with a shunt resistante $R_s$. This means taking
$R_s\rightarrow\infty$, or actually assuming $R_s \gg R_0$ for the
array. The assumptions (i) and (ii) lead  to Eq. (\ref{dtheta})
for $\theta_\mathbf{r}$, which corresponds to the conservation of
supercurrents at each node plus a leakage of normal current to the
substrate. In this case we get
$$\Gamma_\theta =(2e/\hbar)^2 R_0.$$
However, if one considers Eq. (\ref{dtheta}) alone for the
calculation of current-voltage curves with open boundary
conditions it is found that an applied external current leads to
dissipation only at the boundaries were current is injected
(extracted), since normal current will flow directly from the
first (last) row of junctions to the substrate through $R_0$
\cite{tiesinga2}. Strictly periodic boundary conditions are not
possible to be implemented in a consistent way. (iii) In order to
correctly model current-voltage curves and to be able to implement
fluctuating twist periodic boundary conditions, one has to add
\cite{kim99,kim99b} a global normal current channel in parallel to
the whole array, with a "global resistance" $R_\mathrm{global}$,
such that in the normal state the total resistance of  the array
will be given by $R_\mathrm{global}$. Then total conservation of
current leads to Eq. (\ref{dalpha}) which represents a parallel
circuit of the average supercurrent in the array and the global
normal current. In the approach of Refs.
\onlinecite{kim99,jensen,kim99b} $R_\mathrm{global}=R_0$ is
assumed, and therefore this leads to the choice $\Gamma_\alpha =
(2e/\hbar)^2 R_0/L^2$.

The above assumptions (i)-(iii) for the on-site dynamics give a
consistent interpretation of calculations of the current-voltage
response and phase dynamics, but correspond to a model system
rather than to a particular JJA available experimentally. In
realistic JJA the normal currents in the junctions can not be
neglected since usually $R_s\ll R_0$, and therefore the RSJ model
can be a good representation of the JJA. A possible realization of
the dynamics of Eqs. (\ref{dtheta})-(\ref{dalpha}) could be
achieved experimentally if one adds on purpose a resistor in
parallel to the whole array  such that it has a resistance
$R_\mathrm{global} \ll R_s$, in which case normal currents will
mainly go through $R_\mathrm{global}$ and reduce the weight of the
normal quasiparticle currents of the junctions\cite{hernan}.

A good candidate for the on-site dynamics is a superconducting
wire network. In this case one has to take into account the
dynamics of the complex order parameter which is given by the time
dependent Ginzburg-Landau equation (TDGL) coupled to the
electromagnetic field equations \cite{tdgl,tdgl2,tdgl3}.
There are two dissipative mechanisms in this case:
(i) via the normal state
resistivity, since the total current is the sum of the
supercurrent and the normal current in each wire of the network
(this is the equivalent of the shunt resistance of the RSJ model
in a JJA), and (ii) via the relaxation of the complex  order
parameter in the TDGL equations, which is local in nature\cite{tdgl3}
and where its dominant contribution is determined by $D$,
the normal state diffusion constant. After
writing the TDGL equations in a discrete lattice, and neglecting
the fluctuations of the amplitude of the order parameter (London
limit), one obtains\cite{unp} that the on-site part in the
dynamics of the phase is provided by $D$, and would correspond to
Eq. (\ref{dtheta}) with $\Gamma_\theta =  16 \pi^3 D \lambda^2/
(\Phi_0^2 a S)$, where $a$ is the network lattice constant and $S$
the section of the wires. Therefore, there is no need to invoke a
``leakage of normal current to the ground" in this case. The full
dynamics of the superconducting wire network is  a mixture of both
the ``on-site" dynamics and the ``RSJ" dynamics. However, in the
presence of an on-site contribution, the resulting rate of change
of the phases at different sites, like eq. (8), does not have a
logarithmic nonlocal dependence at large separations as in the
pure RSJ model. \cite{shenoy}

In any case, in the present work, we will take the purely on-site
dynamical equations as a model dynamics that corresponds to a
limit of the general dynamics of a JJA or a superconducting wire
network where only local dissipation is taken into account. The
opposite limit for the dynamics is the pure RSJ model that we have
analyzed in Ref. \onlinecite{gd03}.

We integrate the dynamical equations with a second order
Runge-Kutta-Helfand-Greenside method with time step $\Delta
t=0.01-0.07\tau $, averaging over, typically,  $10^6$ time steps
after using $5 \times 10^5$ time steps  for equilibration. The
results were averaged over $5-10$ different initial configurations
of the phases and system sizes ranging from  $L=8 $ to $L=180$
were considered.

\section{Dynamic scaling theory}

Near a second-order phase transition, the diverging correlation
length $\xi $ leads to critical slowing down characterized by
relaxation times $\tau$ that also diverge approaching the
transition temperature. The dynamic scaling hypothesis \cite{hh}
asserts that measurable quantities should scale with the diverging
correlation length $\xi $ and the relaxation time $\tau \propto
\xi ^{z}$, near the transition temperature, where $z$ is the
dynamical critical exponent. A general dynamic scaling theory for
the resistivity behavior near a superconducting transition has
been provided by Fisher, Fisher, and Huse.\cite{ffh} According to
this scaling theory, the nonlinear resistivity $E/J$ should
satisfy the scaling form
\begin{equation}
T\frac{E}{J}=\xi ^{-z}g_{\pm }(\frac{J}{T}\xi )  \label{scaling}
\end{equation}
in two dimensions, where the $+$ and $-$ correspond to the
behavior above and below the transition, respectively. For a
transition in the KT universality class, the correlation length
should diverge exponentially as $\xi \propto \exp (b
/|T/T_{c}-1|^{1/2})$, above $T_c$. Otherwise, for a usual
continuous transition,  a power-law behavior is
expected, $\xi \propto |T/T_{c}-1|^{-\nu }$, with an exponent $\nu
$ to be determined. Thus, a scaling plot according to Eq.
(\ref{scaling}) can be used to verify the dynamic scaling
hypothesis and the assumption of an underlying equilibrium
transition.

The scaling form of Eq. \ref{scaling} does not take into account
finite-size effects and so it is valid only in a range of
temperature $T$ and current densities $J$ where such effects are
not dominant. Finite-size effects are very important sufficiently
close to the transition when the correlation length $\xi$ reaches
the system size $L$. In particular,  at $T_c$, the correlation
length $\xi$ will be cut off by the system size $L$ in any finite
system. From Eq. (\ref{scaling}), the nonlinear resistivity at
$T_c$ should then satisfy the scaling form
\begin{equation}
T\frac{E}{J}=L^{-z}g(\frac{J}{T}L )  \label{finitesize}
\end{equation}

It follows from Eq. (\ref{finitesize}) that the linear resistance
$R_{L}=\lim_{J\rightarrow 0}E/J$ should decrease as a power-law of
the system size,
\begin{equation}
R_L \propto L^{-z}, \label{Lres}
\end{equation}
right at $T_c$. This behavior is independent of the form of
the correlation length divergence. The linear resistance can be
obtained from the Kubo formula of equilibrium voltage fluctuations
as
\begin{equation}
R_L=\frac{1}{2T} \int d t \langle V(t) V(0) \rangle . \label{kubo}
\end{equation}
without an imposing driving current. $R_L$ can also be determined
more accurately from the long-time fluctuations of the total phase
difference across the system $\Delta \theta(t)=L\alpha_\mu$ as
\cite{eg98,jensen}
\begin{equation}
R_L=\frac{1}{2Tt} \langle (\Delta \theta(t)-\Delta
\theta(0))^2\rangle  ,
\end{equation}
valid for sufficiently long times $t$.

The critical dynamics leading to the resistivity scaling described
above can also be studied by the behavior of time correlation
functions. For the frustrated JJA, there are two different types
of time correlations of particular interest, the time correlation
for chiralities $C_{ch}(t)$ and phase variables $C_{ph}(t)$. We
shall use normalized time correlation functions defined as
\begin{equation}
C(t) = \frac{ \langle A(t) A(0)\rangle  - \langle A\rangle^2 }
{\langle A^2\rangle - \langle A\rangle^2 } \label{corr}
\end{equation}
For the phase variables, $A = \vec S=\sum_i \vec s_i$, where $\vec
s =(\cos(\theta),\sin(\theta))$) and for the chiral variables $A=
\chi = \sum_{<ij>} (\theta_i -\theta_j - A_{ij}) /2\pi$, where the
summation is taken over the elementary plaquette of the lattice
and the gauge-invariant phase difference is restricted to the
interval $[-\pi,\pi]$. The relaxation time $\tau $  can be
obtained from the exponential decay $C(t) \propto \exp(-t/\tau)$
at sufficiently long times. In general, the time dependence of
$C(t)$ can be expressed as a series of exponential terms with the
largest decay time corresponding to the critical relaxation time
of the long time dynamics.\cite{landau}. From dynamic finite-size
scaling, the relaxation time should scale at $T_c$ as $\tau
\propto L^z$, from which the $z$ can be estimated from the slope
in a log log plot. An alternative procedure to estimate $z$ from
equilibrium dynamics is to explore the expected finite-size
behavior of the time correlation functions at long times. Since at
$T_c$ the relaxation time scales as $\tau \propto L^z$, the time
correlation function for different system sizes can be cast into a
scaling form in terms of the dimensionless ratio $t/L^z$ as
\begin{equation}
C(L,t)=\tilde C(t/L^z)                \label{corrscal}
\end{equation}
where $\tilde C(x)$ is a scaling function. However, this assumes a
simple scaling form for the time correlation functions and is only
valid for sufficiently long times when a single exponential term
describes the relaxation behavior.

\section{Results and scaling analysis}

\begin{figure}
\includegraphics[bb= 1cm  2.5cm  19cm   26.5cm, width=7.5 cm]{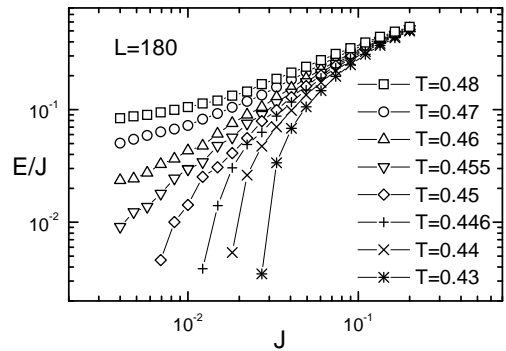}
\caption{ Nonlinear resistivity $E/J$ as a function of temperature
for system size $L=180$.} \label{resl128}
\end{figure}

Fig. \ref{resl128} shows the temperature dependence of the
nonlinear resistivity $E/J$ for the largest systems size $L=180$
near the chiral transition temperature $T_{ch}$, estimated
previously from equilibrium Monte Carlo simulation,\cite{gkn}
$T_{ch}=0.455$. Qualitatively, the linear resistance
$R_{L}=\lim_{J\rightarrow 0}E/J$, tends to a finite value at high
temperatures but extrapolates to very low values at lower
temperatures, consistent with the existence of a resistive
transition  in the range $0.45 < T_{c} <0.46 $. In the double
transition scenario, where the phase-coherence transition is
expected to be in the KT universality class, the estimate of the
proposed KT critical temperature is $T_{KT}=0.446$, from Monte
Carlo simulations,\cite{olsson} which is close to $T_{ch}$.
However, as it is clear from the behavior at the lowest currents
in Fig. \ref{resl128},  this estimate is below the resistive
transition since the resistivity curves for $T=T_{KT}=0.446$ and
$T=0.45 > T_{KT}$ tends to zero for $J\rightarrow 0$, indicating
that the system is still in the superconducting phase for these
temperatures. On the other hand, the resistivity curve for $T=0.46
> T_{ch}$ clearly tends to finite resistivity for $J\rightarrow
0$. This shows that the resistivity transition occurs at $T_{ch}$
or at a temperature very close to $T_{ch}$ rather than at the
proposed estimate of $T_{KT}$.

\begin{figure}
\includegraphics[bb= 1cm  2.5cm  19cm   26.5cm, width=7.5 cm]{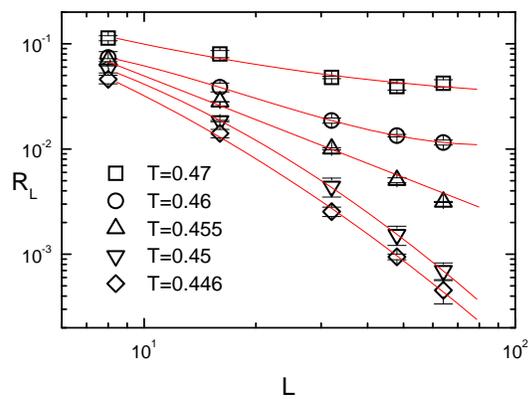}
\caption{ Linear resistance $R_L$, obtained without current bias,
as a function of temperature and system size. Lines are just guide
to the eyes} \label{lresxl}
\end{figure}

Additional support for a resistivity transition at $T_{ch}=0.455$
is provided by the behavior of the linear resistivity $R_L$ as a
function of system size, shown in Fig. \ref{lresxl}. For $T >
0.455$, $R_L$ extrapolates to a finite value consistent with the
behavior of the nonlinear resistivity for $J\rightarrow 0 $ in
Fig. \ref{resl128}. On the other hand, for $T \le 0.455$ it
extrapolates to zero, indicating that the resistive transition
temperature is compatible with the estimate of $T_{ch}=0.455$.
Since in this calculations $R_L$ is obtained without any current
bias, from the equilibrium dynamical fluctuations, according to
Eq. \ref{kubo}, this result also verify that the $T_c$ inferred
from the behavior of the nonlinear resistivity for the largest
system size in Fig. \ref{resl128} is not an artifact of finite
current bias and in fact reflects the underlying equilibrium
transition behavior.

\begin{figure}
\includegraphics[bb= 1cm  2.5cm  19cm   26.5cm, width=7.5 cm]{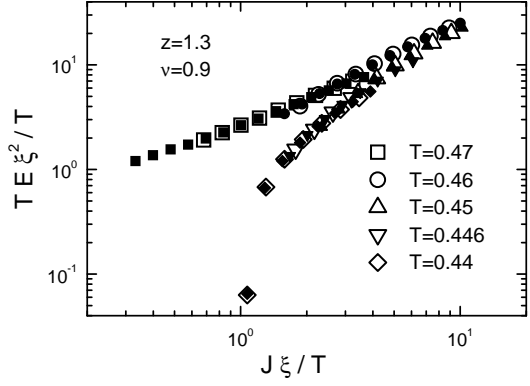}
\caption{Scaling plot of the nonlinear resistivity data for the
smallest current densities near $T_c=T_{ch}=0.455$ with $\xi
\propto |T/T_{c}-1|^{-\nu }$. Open symbols correspond to $L=128$
and filled ones to $L=180$. } \label{rscalT}
\end{figure}

\begin{figure}
\includegraphics[bb= 1cm  2.5cm  19cm   26.5cm, width=7.5 cm]{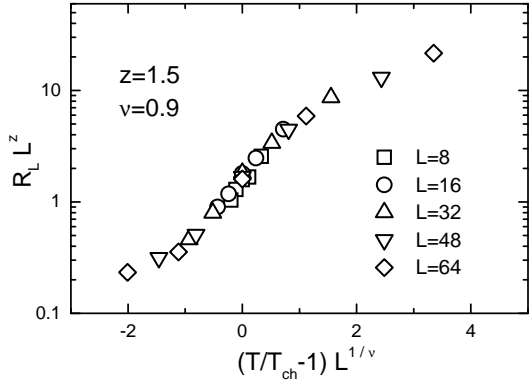}
\caption{Finite-size scaling plot of the linear resistance data
near $T_c=T_{ch}=0.455$.} \label{lrescal}
\end{figure}

Although the resistivity behavior of Figs. \ref{resl128} and
\ref{lresxl} already suggest that the resistive transition
temperature coincides with $T_{ch}$ or it is much  closer to this
value than previous estimates, we now proceed, as in any study of
critical phenomena, to  obtain the asymptotic equilibrium critical
behavior in the thermodynamic limit,  $L \rightarrow \infty $ and
$J\rightarrow 0$, from a scaling theory.  A scaling plot according
to Eq. \ref{scaling} is shown in Fig. \ref{rscalT} for the largest
system sizes, in the temperature range closest to $T_{ch}$ and
smallest current densities, assuming the correlation length $\xi $
has a power-law divergence with $T_{c}=T_{ch}$ and using $\nu$ and
$z$ as adjustable parameters so that the best data collapse is
obtained. This scaling plot shows that the two largest system
sizes $L=128$ and $L=180$ give the same data collapse and so
finite size effects neglected in the scaling form of Eq.
\ref{scaling} are not dominant for the range of temperatures and
current densities shown in the plot. Similar scaling analysis
assuming a KT correlation length and fixing $T_c$ at the estimate
of $T_{KT}$ does not result in a good data collapse. The same
behavior was found using the RSJ dynamics.\cite{gd03} From this
scaling analysis, we estimate $\nu=0.9(1)$ and the dynamical
critical exponent $z=1.3(3)$. The static exponent $\nu$ is
consistent with estimates of the chiral transition from
equilibrium Monte Carlo simulations \cite{gkn} but the accuracy is
not sufficient to rule out the value $\nu=1$ expected for the
standard Ising transition. Our estimate of $z$ is smaller than the
one obtained previously for the frustrated XY model with MC vortex
dynamics \cite{leeteitel} where $z\sim 2$ was found. However, such
MC simulation corresponds to a different dynamics and also only
very small system sizes (with $L=8-14$) were analyzed. We now take
into account finite-size effects explicitly by studying the
scaling behavior of the linear resistivity $R_L$ near $T_c$ in
Fig. \ref{lresxl}. At $T_c$, the linear resistivity should scale
with system size  according to Eq. \ref{Lres}. Near $T_c$, it
should also depend on temperature through the dimensionless
variable $L/\xi$. If the correlation length diverges as a power
law then it should satisfy the finite-size scaling form
\begin{equation}
R_L L^z = f((T/T_c -1)L^{1/\nu})   \label{lrest}
\end{equation}
In fact, as shown in Fig. \ref{lrescal} the linear resistivity
data satisfy the scaling form with $T_c=T_{ch}$ and a value
$z=1.5(2)$ consistent with the estimate from the nonlinear
resistivity scaling.

The above scaling analysis for the nonlinear resistivity at large
system sizes and linear resistivity at smaller system sizes
already confirm that the resistive transition temperature $T_c$ is
very close to $T_{ch}$,  with a dynamic exponent $z < 2$. However,
in the absence of a completely satisfactorily determination of
$T_c$ from static critical behavior, \cite{rj,gkn96,olsson} from
now on, we will assume $T_c=T_{ch}$ and explore to which extent
this give us consistent results for the dynamical critical
behavior, including finite-size effects. Another reason to assume
the value of $T_c$ obtained from equilibrium simulations rather
than estimating from the dynamic scaling itself is that, in
general, the most reliable way of studying critical dynamics and
determine the dynamic exponent $z$ is to use the known value of
$T_c$. This is true not only for models where $T_c$ is known
exactly as for the two-dimensional Ising model \cite{grant} but
also for models where $T_c$ is only known by numerical simulations
as for the three-dimensional Ising model. \cite{landau}

\begin{figure}
\includegraphics[bb= 1cm  2.5cm  19cm   26.5cm, width=7.5 cm]{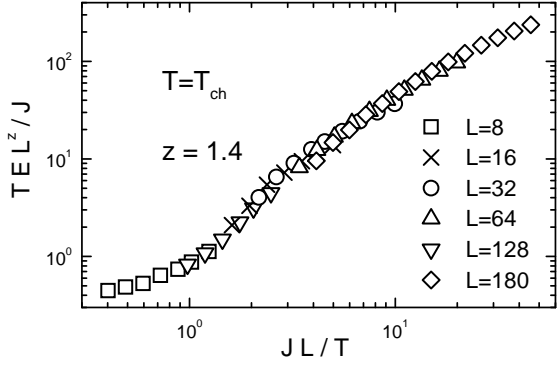}
\caption{Scaling plot of the nonlinear resistivity $E/J$ at
$T_c=T_{ch}=0.455$ for different system sizes $L$. }
\label{rscalTch}
\end{figure}

\begin{figure}
\includegraphics[bb= 1cm  2.5cm  19cm   26.5cm, width=7.5 cm]{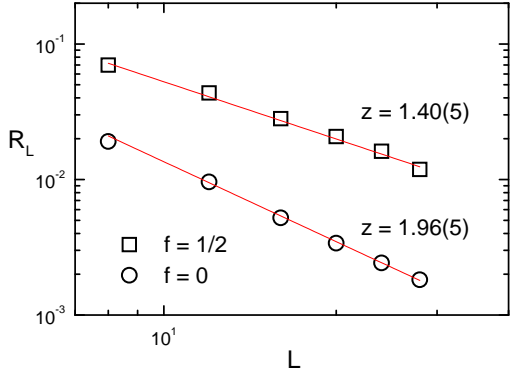}
\caption{Linear resistance as a  function of system size at the
critical temperatures $T_c=T_{ch}$ for $f=1/2$ and $T_c=0.887$ for
$f=0$. Power-law fits give estimates of the dynamic exponent $z$.
} \label{LresTch}
\end{figure}

An alternative estimate of $z$ can be obtained from the nonlinear
resistivity by studying the expected size dependence at $T_c$. As
shown in Fig. \ref{rscalTch}, a finite size scaling according to
Eq. (\ref{finitesize}) gives  the same dynamic exponent
$z=1.4(3)$, within the estimated error bar. The same behavior was
also observed using the RSJ dynamics \cite{gd03} but with a
smaller value of $z$. Equilibrium calculations of the linear
resistance $R_L$ at $T_{ch}$ also give a consistent estimate. Fig.
\ref{LresTch} shows the finite size behavior of $R_L$ obtained
from Eq. (\ref{Lres}). A power-law fit gives $z=1.41(5)$ which
agrees with the other estimates  and suggests therefore that the
value of $z$ corresponds to the underlying equilibrium dynamical
behavior. To show the reliability of this method, it is also
included in Fig. \ref{LresTch} the behavior for the unfrustrated
case, $f=0$. In this case the resistive transition is in the KT
universality class and  a dynamical exponent $z=2$ is expected,
independent of the dynamics. Indeed, for $f=0$, the same power-law
fit at the critical temperature $T_c=0.887$ estimated from Monte
Carlo simulations \cite{weber} gives $z=1.96(5)$, in good
agreement with previous resistivity calculations \cite{jensen} for
$f=0$ using smaller system sizes up to $L=16$.

It should be noted that our above estimate of the dynamic exponent
$z$ is obtained by requiring that $T_c$, $z$ and $\nu$ satisfy at
the same time the finite-size scaling forms of  Eqs.
(\ref{finitesize}), (\ref{Lres}) and (\ref{lrest}), including
small system sizes, as well as the scaling form of Eq.
(\ref{scaling}) for the largest system sizes. Using only Eq.
(\ref{scaling}) can lead to inaccurate estimates of $z$ as shown
recently in Ref. \onlinecite{holzer} for the unfrustrated case.

\begin{figure}
\includegraphics[bb= 1cm  2.5cm  19cm   26.5cm, width=7.5 cm]{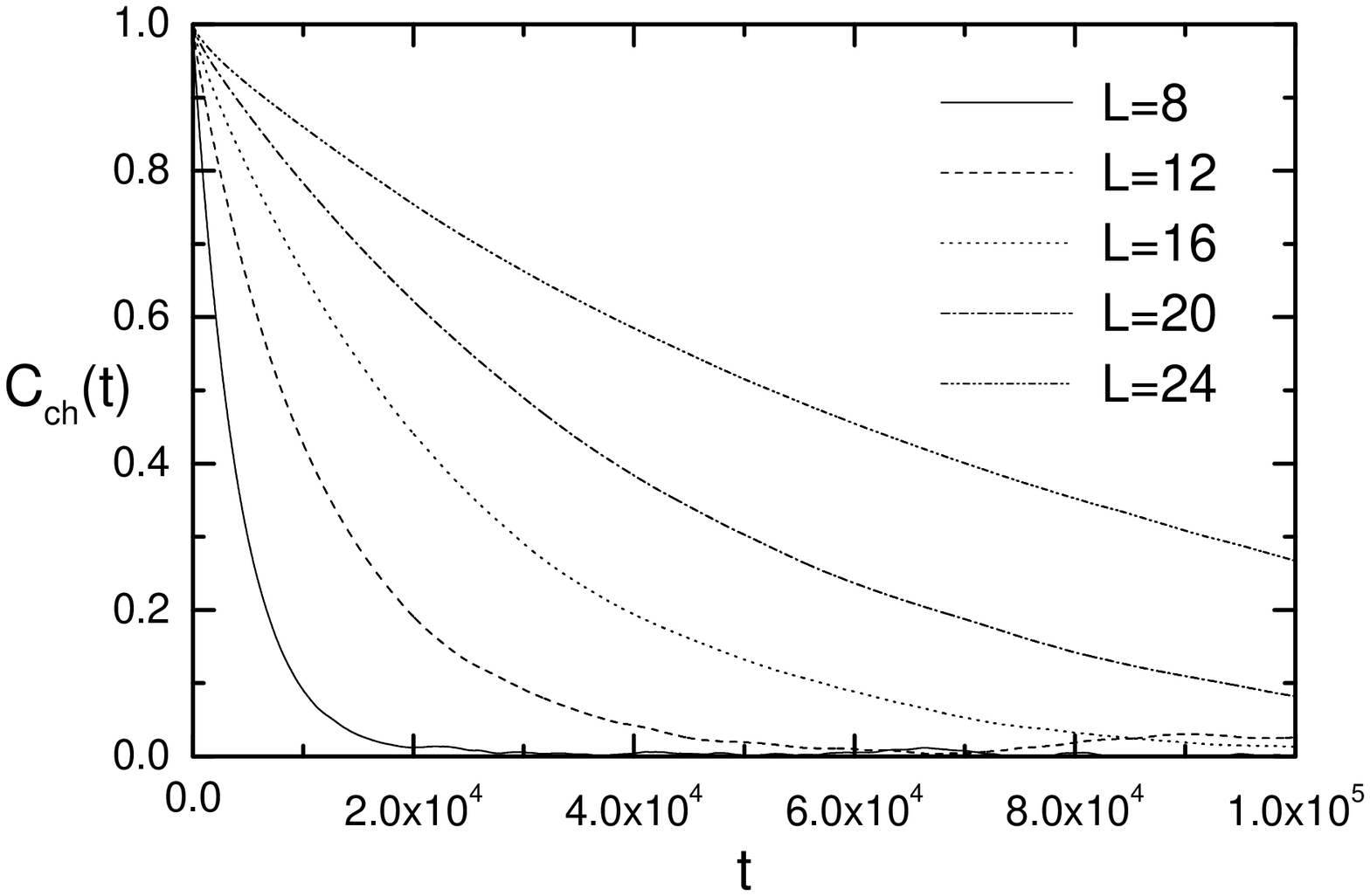}
\caption{Time correlation function $C_{ch}(t)$ for the chiral
variables at $T_{ch}$, for different system sizes. }
\label{corrch}
\end{figure}

\begin{figure}
\includegraphics[bb= 1cm  2.5cm  19cm   26.5cm, width=7.5 cm]{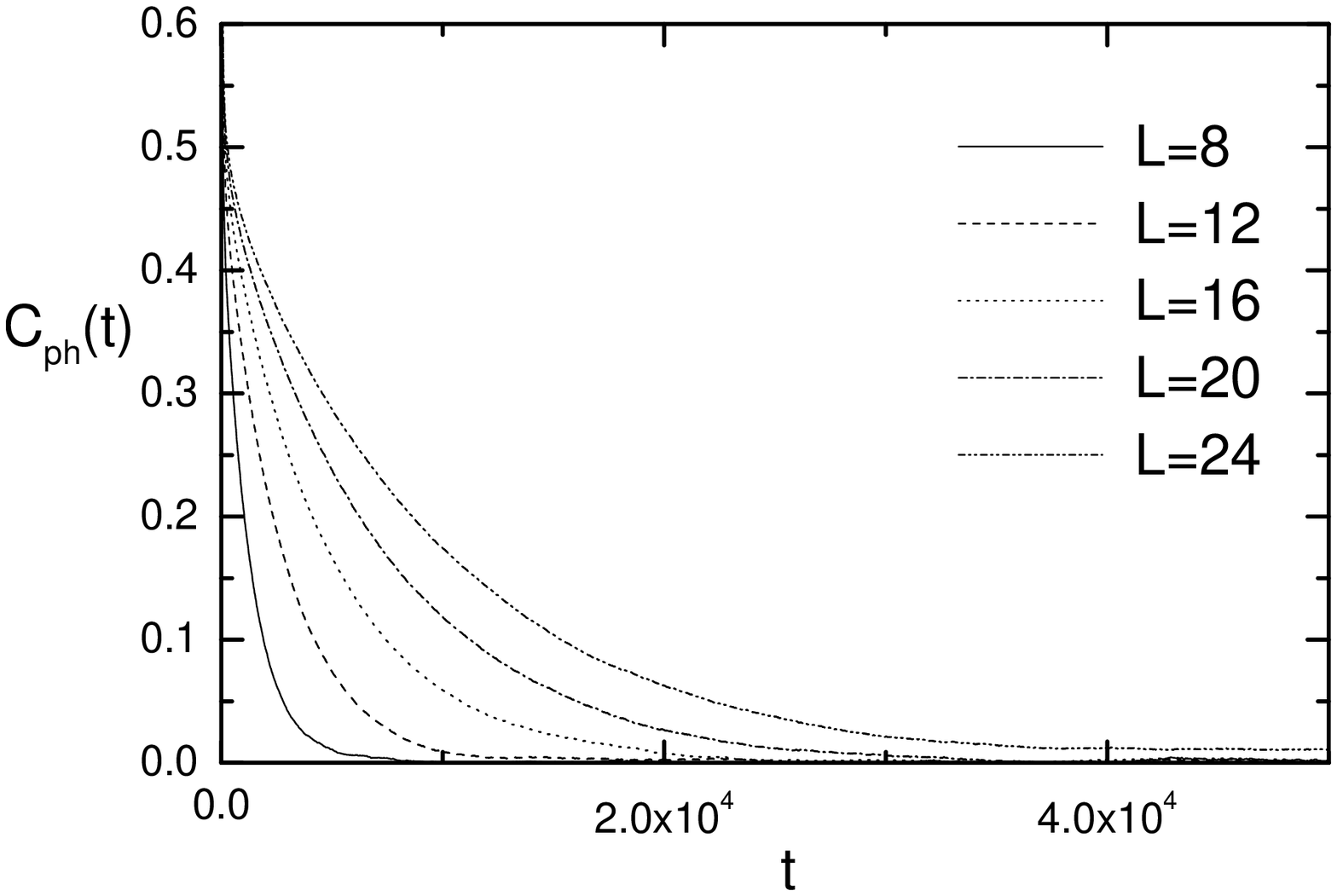}
\caption{Time correlation function $C_{ph}(t)$ for the phase
variables at $T_{ch}$, for different system sizes. }
\label{corrph}
\end{figure}

\begin{figure}
\includegraphics[bb= 1cm  2.5cm  19cm   26.5cm, width=7.5 cm]{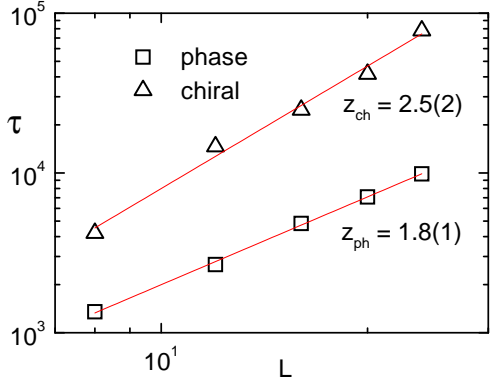}
\caption{Finite-size behavior of the phase and chiral relaxation
times,  $\tau_{ph}$ and $\tau_{ch}$ respectively,  at the critical
temperature $T_c=T_{ch}$. Power-law fits give estimates of the
dynamical exponents $z_{ph}$ and $z_{ch}$. } \label{tauxL}
\end{figure}

\begin{figure}
\includegraphics[bb= 1cm  2.5cm  19cm   26.5cm, width=7.5 cm]{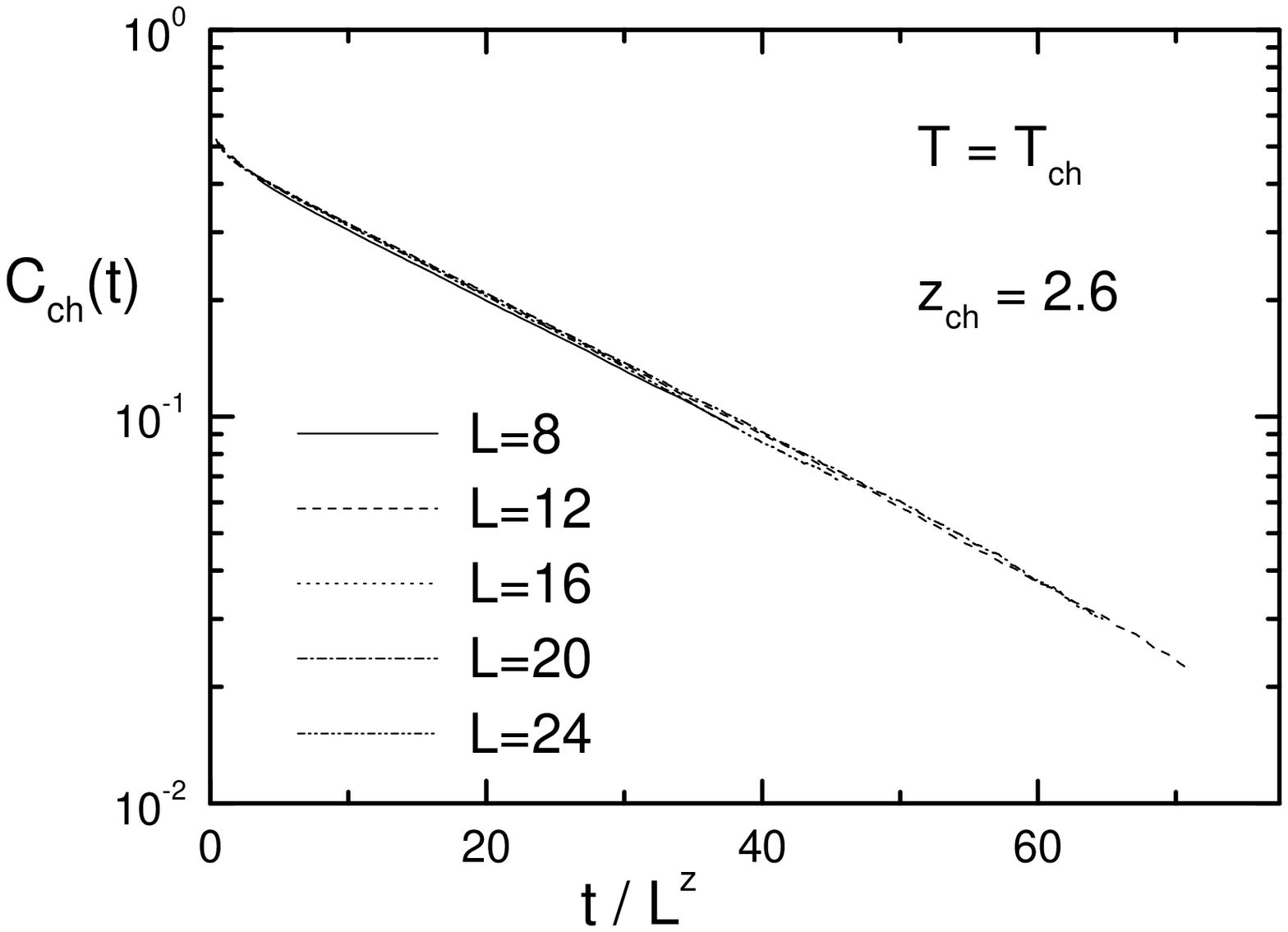}
\caption{Finite-size scaling plot of the time correlation function
$C_{ch}(t)$. } \label{cscalch}
\end{figure}

\begin{figure}
\includegraphics[bb= 1cm  2.5cm  19cm   26.5cm, width=7.5 cm]{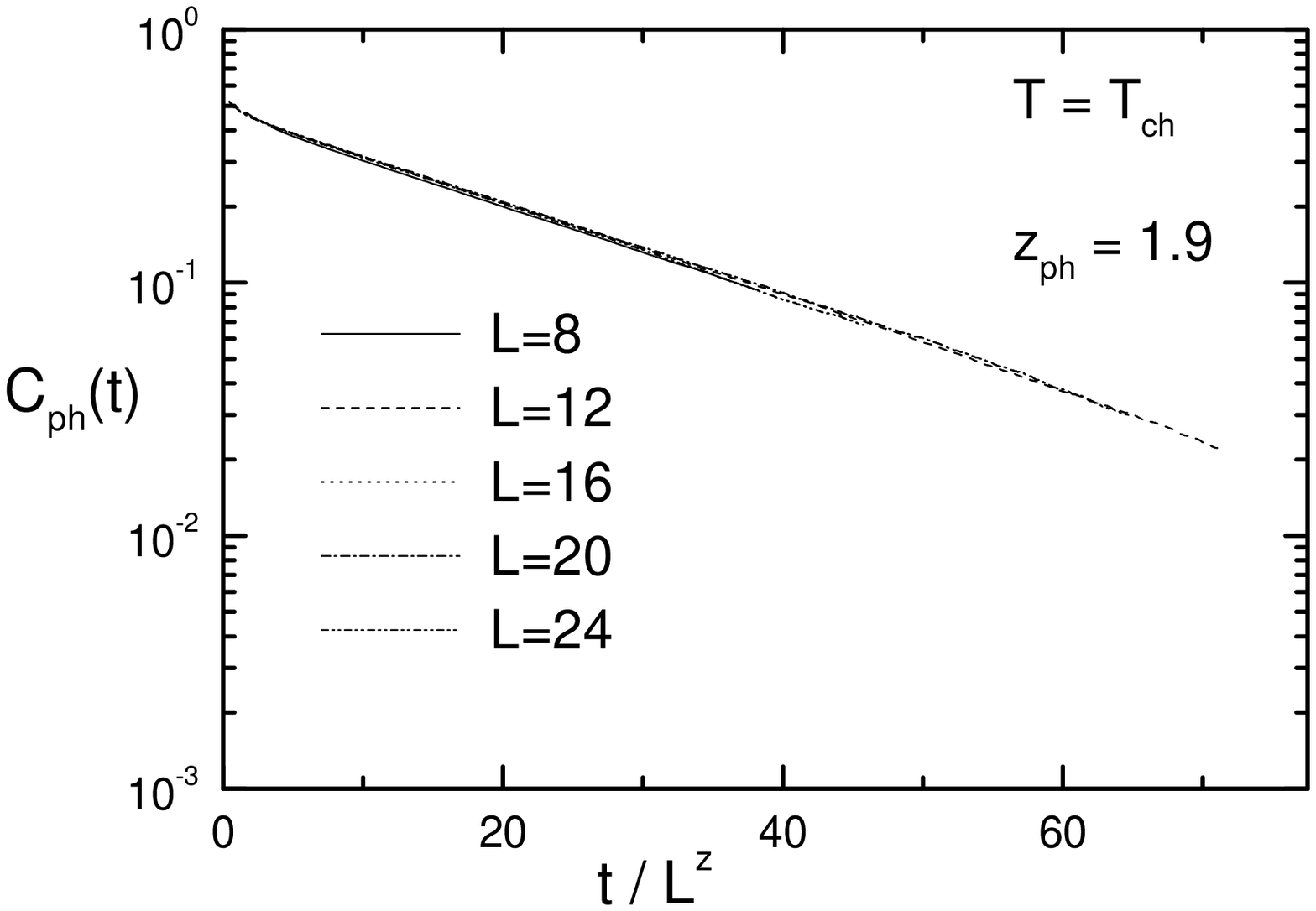}
\caption{Finite-size scaling plot of the time correlation function
$C_{ph}(t)$. } \label{cscalph}
\end{figure}

\begin{figure}
\includegraphics[bb= 1cm  2.5cm  19cm   26.5cm, width=7.5 cm]{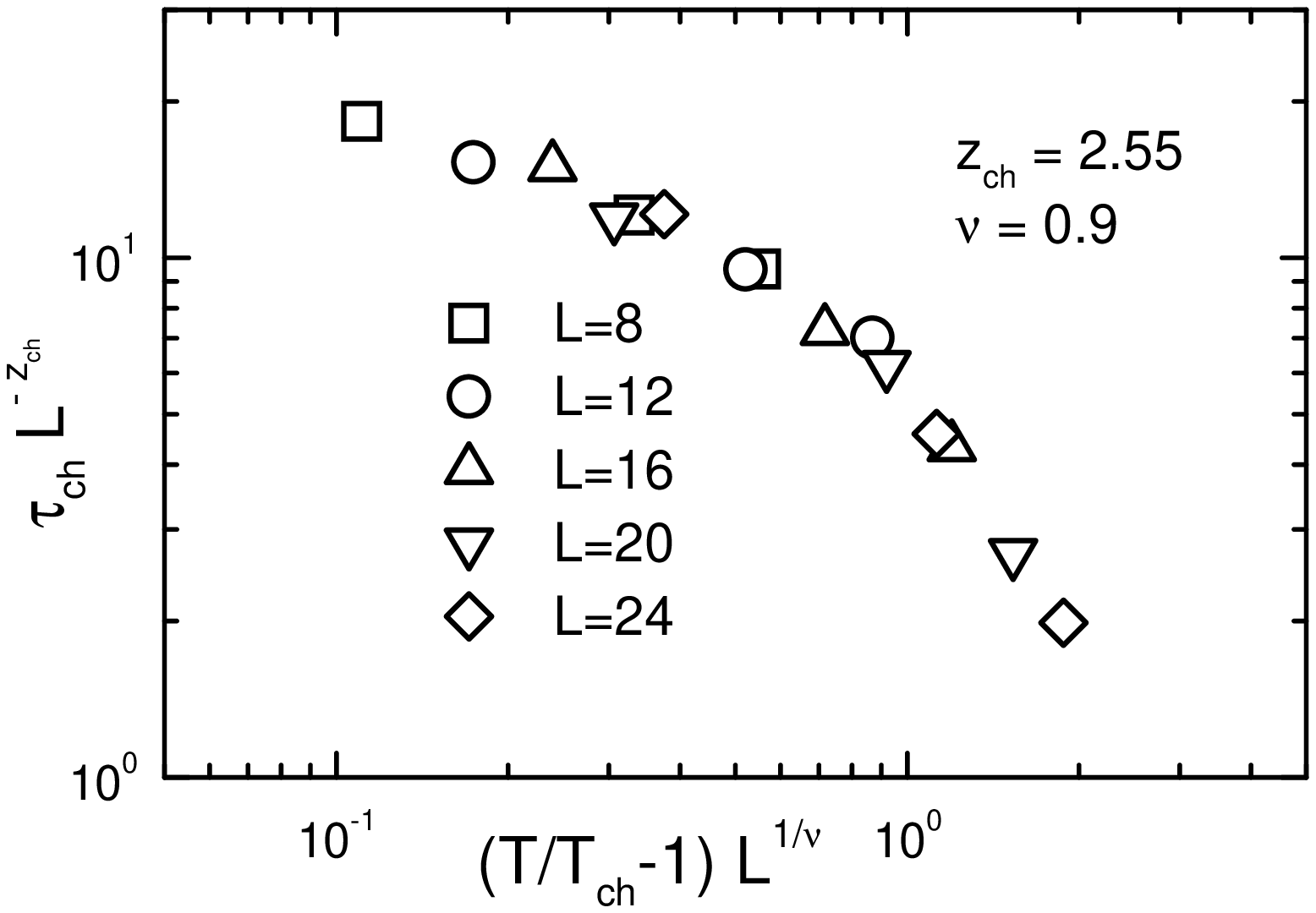}
\caption{Finite-size scaling plot for the relaxation time
$\tau_{ch}$ } \label{chtauxT}
\end{figure}

\begin{figure}
\includegraphics[bb= 1cm  2.5cm  19cm   26.5cm, width=7.5 cm]{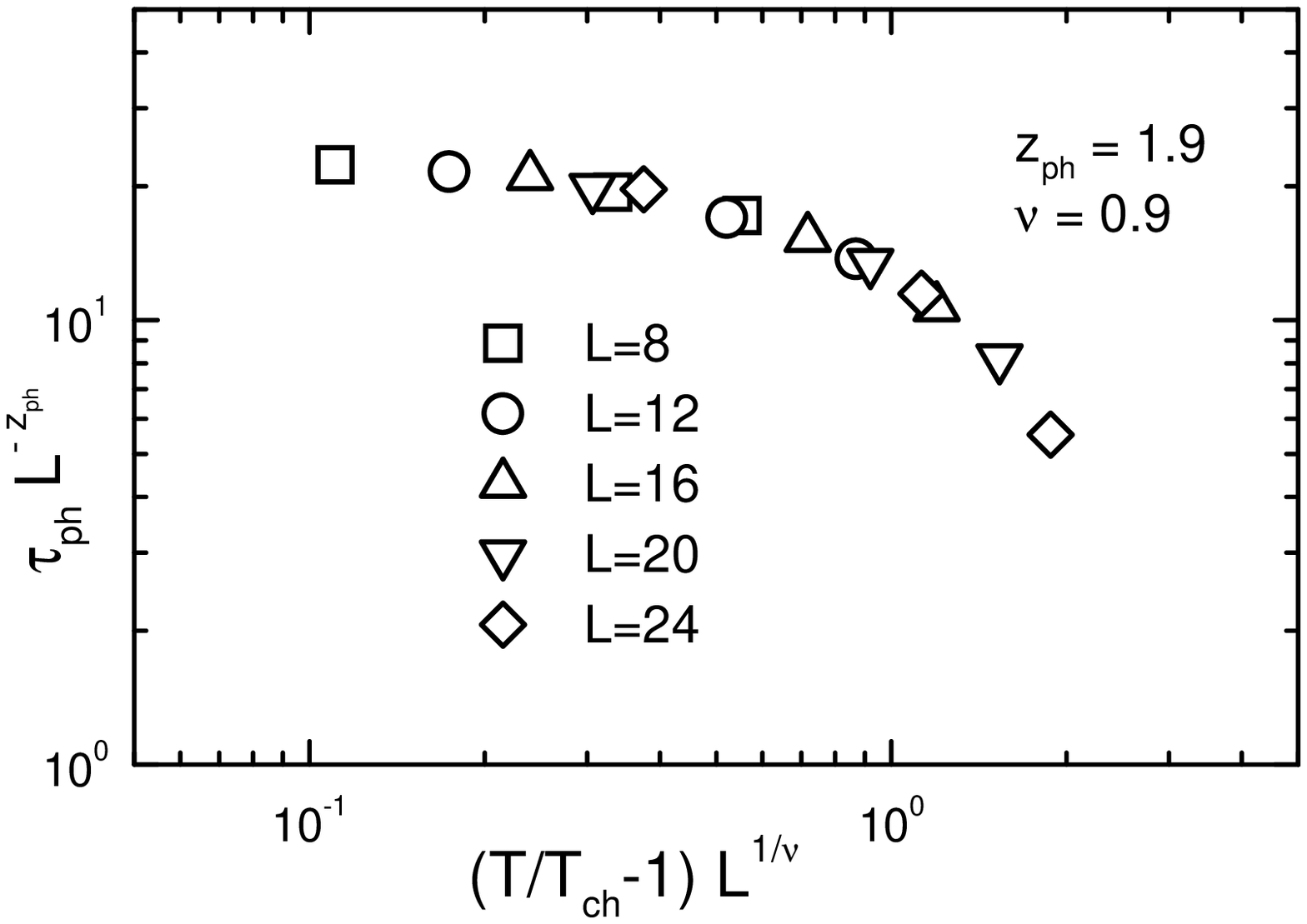}
\caption{Finite-size scaling plot for the relaxation time
$\tau_{ph}$ } \label{phtauxT}
\end{figure}

To further verify that the estimate of $z$ obtained from the
resistivity scaling does in fact reflect critical phase
fluctuations near the transition rather than just critical
fluctuations for the chiral order parameter, we have also
performed equilibrium calculations of the phase autocorrelation
functions $C_{ph}(t)$ for the phase variables and $C_{ch}(t)$ for
the chirality variables. Figs. \ref{corrch} and \ref{corrph} show
the finite-size behavior of the time correlations functions
evaluated at the critical temperature $T_{ch}$. If this
temperature corresponds to the critical point for phase coherence
and vortex-lattice disorder then the relaxation times for both
phase and chirality variables should diverge with the system size
as $\tau \propto L^z$. The relaxation time $\tau_{ph}$ and
$\tau_{ch}$ can be obtained from the exponential decay of $C(t)$
at sufficiently long times. We take into account possible
contributions from short time behavior  by fitting the time
dependence of $C_{ph}(t)$ and $C_{ch}(t)$ to a sum of two
exponentials and extract $\tau$  from the largest decay time. Fig.
\ref{tauxL} shows the finite-size behavior of the relaxation time
at $T_{ch}$ for the phases and chiralities. From a power-law fit
we obtain $z_{ph}=1.8(1)$ from the phase relaxation time
$\tau_{ph}$ which is indeed consistent, within the estimated error
bar, with the value of $z$ obtained from the resistivity scaling
discussed above. The estimate from the chiral relaxation time in
Fig. \ref{tauxL} is significantly different, $z_{ch}=2.5(2)$.  For
an alternative estimate of $z$ we have also used the scaling of
the correlation function itself. The time correlation functions
should satisfy the scaling behavior of Eq. \ref{corrscal}. As
shown in Figs. \ref{cscalch} and \ref{cscalph}, $C_{ph}(t)$ and
$C_{ch}(t)$ indeed satisfy the expected finite size behavior at
the critical temperature providing additional estimates of the
dynamic exponents $z_{ph}=1.9(2)$ and $z_{ch}=2.6(2)$ which are
consistent, within the estimated error bar, with the values
obtained from the relaxation time scaling. Finally, above $T_{c}$,
the relaxation time should depend both on system size and
temperature. If the correlation length diverges as a power law
then $\tau_{ph}$ and $\tau_{ch}$ should satisfy the finite-size
scaling form
\begin{equation}
\tau L^{-z} = f((T/T_{c} -1) L^{1/\nu} )
\end{equation}
In fact, the data collapse in Figs. \ref{chtauxT} and
\ref{phtauxT} show that this scaling form is satisfied with
$T_c=T_{ch}$ and the values of $\tau_{ph}$ and $\tau_{ch}$ which
are consistent with the above estimates.

\section{Discussion}

Recently, Holzer {\it et al.} \cite{holzer} showed that for the
unfrustrated case, $f=0$, the scaling behavior in Eq.
(\ref{scaling}) considered alone, i.e., without taking into
account finite-size effects, yields incorrect values for the
dynamic exponent $z$, using approximate analytical expressions for
the IV characteristics available in the literature \cite{conf}. We
should emphasize that our approach for the resistivity scaling
analysis described in the previous Section is quite different.
Our estimate of the dynamic exponent $z$ is obtained by requiring
that $T_c$, $z$ and $\nu$ satisfy at the same time the finite-size
scaling forms of Eqs. (\ref{finitesize}), (\ref{Lres}) and
(\ref{lrest}), including small system sizes, as well as the
scaling form of Eq. (\ref{scaling}) for the largest system sizes.
It should also be considered that the possibility of an
equilibrium KT transition for $f=1/2$ within the separated
transitions scenario does not imply that the dynamics would be the
same as the KT dynamics and therefore for the frustrated case
considered here there is no reliable analytical expressions
available for the IV characteristics. The dynamics for $f=1/2$
will be different because besides vortex excitations, chiral
domain walls also contribute to the nonlinear resistivity as shown
in Ref. \onlinecite{teitel89}. Moreover, it has already been shown
for the $f=0$ case that, when finite-size scaling is taking into
account in the resistivity scaling theory of Fisher {\it et al.}
\cite{ffh}, as we also do in our approach, the correct dynamic
exponent $z=2$ is obtained for the KT transition, as shown for
example in Ref. \onlinecite{jensen}. This is also verified in the
scaling analysis of our data as shown in Fig. \ref{LresTch}, where
we find a dynamic exponent consistent with $z=2$ for $f=0$, as
expected.

The distinct values obtained for $z_{ph}$ and $z_{ch}$ with the
on-site dissipation model deserve some considerations. Similar
behavior was also found by us using the RSJ dynamics.\cite{gd03}
The final results for both models, obtained from the resistivity
scaling and time-correlation function scaling analysis, are
summarized in Table 1. Although for the on-site model, the two
methods give results for $z_{ph}$ which differ beyond the
estimated errorbar, the values are significant below the value
obtained for $z_{ch}$. Naively, if the two transitions happen at
the same temperature, one would expect that the same dynamic
exponent should hold for the phase and chiral relaxation times.
However, we should mention that different dynamic exponents for
coupled order parameters have already been found previously at
multicritical points in magnetic systems \cite{huber}. This
suggests that a possible explanation for two dynamic exponents at
the transition of the FJJA may rely on the existence of a
multicritical point in the phase diagram of the relevant effective
Ginzburg-Landau free energy describing the transition. A
multicritical point is known to occur in the coupled XY-Ising
model \cite{gkn} which should  describe the static critical
behavior of the FJJA and this could be a useful framework for
investigations of the dynamical universality class of FJJA. In the
context of superconducting systems, different dynamic exponents
for the resistivity and chirality have also recently been found in
the resistive transition of disordered superconductors \cite{eg04}
described by the three-dimensional XY spin glass model
\cite{kawamura}. Just as in the case of the frustrated JJA, the
phase transition in the XY spin glass results from the competition
of a chiral order parameter and phase variables. Although earlier
work for this problem concluded for a spin-chirality decoupling
picture \cite{kawamura}, more recent numerical work have provided
strong evidence \cite{young} that there is a single transition at
which both phase variables and chiralities order.

Although the single transition scenario provides a consistent
interpretation of our data, it is worth emphasizing that the
alternative separated transitions scenario \cite{korshu} can not
be ruled out. We believe, there are two possible explanations for
some of our findings within the later scenario, as discussed
below.

It is possible that the  KT transition is actually much closer to
$T_{ch}$ than estimated previously and so the transitions can not
be resolved within the accuracy of our data. Our analysis of the
resistivity behavior suggests that in this case it should occur
above $T_c \sim 0.452$. This value is already  close or within the
range of the errorbars reported for the chiral transition critical
temperature obtained, for example, by Monte Carlo simulations
which gives $T_{ch}=0.455(2)$ ( Ref. \onlinecite{lkg}) or $
0.454(2)$ ( Ref. \onlinecite{leelee}). It should be noted however
that this only considers the critical temperatures alone and not
the critical behavior. In the alternative decoupled single
transition scenario, the critical behavior should be described by
a superposition of a pure KT and pure Ising transitions at the
same critical temperature. However, this is also not consistent
with our results. Nevertheless, even if the transitions are so
close that their critical temperatures can not be resolved by any
method, in principle, it could still be possible to distinguish
these scenarios due to the mechanism discussed in Ref.
\onlinecite{korshu} or due to the effects of different corrections
to scaling.

A second possibility is that the dynamic scaling theory of Fisher
{\it et al.} \cite{ffh} in its original form in Eq. \ref{scaling}
is not valid for the present case and should  be enlarged to
include the interplay of two divergent length scales at nearby
temperatures \cite{olsson} which can lead to crossover effects at
small length and time scales. In fact, the underlying assumption
in the resistivity scaling theory is that there is a single
divergent length scale, corresponding to the leading divergent
contribution to finite correlation lengths, when approaching the
critical temperature of the resistive transition. This would
certainly be valid within the coupled single transition scenario,
which is consistent with our conclusions since in that case phase
coherence and chiral order develop at the same critical
temperature, with strongly coupled order parameters, and the
equilibrium critical behavior should be described by a single
divergent length scale. Above the transition, in the disordered
phase, the chiral and phase correlation lengths diverge when
approaching $T_c$ with a common leading divergent contribution.
Below the transition, where there is chiral order and a Gaussian
fixed line is expected for the phase variables, the chiral
correlation length diverges when approaching $T_c$ with the same
leading divergent contribution while the phase correlation length
remains infinite since the Gaussian fixed line corresponds to the
absence of a length scale.  However, if phase coherence and chiral
order develop at different temperatures then the resistivity
scaling can only hold sufficiently close to the phase coherence
transition otherwise the scaling form of Eq. \ref{scaling} should
be enlarged to include the divergent chiral correlation length in
addition to the phase correlation length. This would lead to a
scaling function $g_{\pm}(x,y)$ in Eq. \ref{scaling} depending on
$2$ scaling variables $x=J \xi_{KT}/T$ and $y=\xi_{ch}/\xi_{KT}$,
which makes the scaling analysis of the data very complicated
specially when taking into account finite-size effects.  This
could explain, for example, why a good scaling collapse like Fig.
\ref{rscalTch} is not obtained by assuming a resistive transition
at $T_c=T_{KT}$, estimated by previous works. However, it would
remain unclear to us in this case why the linear and nonlinear
resistivity scaling as well  as the critical dynamics including
different temperatures and system sizes are so well described by a
resistive transition at $T_c=T_{ch}$.

\section{Conclusions}

We have studied the resistivity scaling and critical dynamics of a
frustrated Josephson-junction array, at $f=1/2$ flux quantum per
plaquette, by numerical simulations of an on-site dissipation
model for the array dynamics. Using a dynamic scaling analysis, we
find that the resistivity behavior and critical dynamics are well
described by the critical temperature corresponding to the chiral
(vortex-lattice) transition with a correlation length that
diverges as a power law. Two dynamic exponents, $z_{ph}\sim 1.5$
and $z_{ch}\sim 2.5$, are found for phase-coherence and chiral
order, respectively. Consequently, at the transition, the exponent
of the current-voltage power-law, $V\sim I^a$, is $a = z_{ph}+1
\approx 2.5$ rather than $a=3$ as for the unfrustrated case. The
same behavior has been found recently for the
resistively-shunted-junction model \cite{gd03} but with different
values for the dynamic exponents ( $z_{ph}\sim 0.9(1)$ and
$z_{ch}\sim 2.1$). One implication of these results for transport
experiments is that the usual method of locating the critical
temperature from the value corresponding to a nonlinear IV
exponent $a=3$, may lead to a significant underestimate. This is
more severe for tunnel-junction arrays which should be better
described by the resistively-shunted-junction model, \cite{shenoy}
where we expect $a\sim 2$ at the resistive transition.
\cite{comment} For wire networks the on-site dissipation model
should be more appropriate. Indeed, resistivity scaling of
experimental data on wire networks \cite{ling} find $z \sim 2$,
which is consistent with our estimate of $z_{ph}$ within errors.
It also shows that the resistivity scaling is well described by a
power-law correlation length as found in our simulations. Further
detailed IV measurements combined with magnetic properties, which
could in principle probe the chiral transition, are needed to test
our results.

\acknowledgments

This work was supported by a joint grant CNPq/Prosul-Brasil (no.
490096/03-4) and in part by FAPESP (Grant no. 03/00541-0) (E.G.)
and CONICET, CNEA and ANPCyT (PICT99 03-06343) (D.D.).



\newpage

\begin{table}
\caption{Dynamic exponents of the resistive chiral transition at
$T_{ch}$ using the on-site dissipation model (TDGL) and
resistively-shunted-junction model (RSJ). The superscripts $R$ and
$C$ correspond to results obtained from the resistivity scaling
and time-correlation function scaling, respectively.}
\begin{ruledtabular}
\begin{tabular}{llll}
 \    & \ \ \ \ \ \ RSJ & \ \ \ \ \ \ TDGL  \\
  \hline
 $z_{ph}$  & 1.1(1)$^C$, 0.9(1)$^R$ & 1.8(1)$^C$, 1.4(1)$^R$  \\
 $z_{ch}$  & 2.1(1)$^C$ & 2.5(2)$^C$  \\
\end{tabular}
\end{ruledtabular}
\end{table}

\begin{references}
\bibitem{tdgl3} The TDGL equations were proposed by Schmid in
Ref. \onlinecite{tdgl}(a) and were shown rigorously to be valid
for gapless superconductors (due to paramagnetic impurities) at
all temperatures in Ref. \onlinecite{tdgl}(b). More general
versions of the TDGL equations, which add a nonlinear term in the
time-derivative,  were obtained in Ref. \onlinecite{tdgl}(c),(d)
valid for dirty superconductors near $T_c$, including several pair
breaking mechanisms, like electron-phonon scattering, paramagnetic
impurities, etc. Comparison with experimental results of the TDGL
equations\cite{tdgl2} in the case of superconducting wires show
that the extra nonlinear terms in the time-derivative are
important for high currents. In general, there is no universal
simple TDGL equation appropriate to all situations in any
material. However, in any case, the relaxation of the
superconducting order parameter is expected to be always local in
space and thus it will provide an on-site dissipation mechanism
which is completely different from the pure RSJ model dynamics. In
this respect the simple TDGL equations of Ref.
\onlinecite{tdgl}(a) seem to be enough to describe the asymptotic
critical dynamics, as well as the behavior for low currents near
the superconducting $T_c$.
\bibitem{comment}The experiments in {\sl underdamped} JJA of Ref.~\onlinecite{zant} were
interpreted using the KT theory obtaining $a\gtrsim 3$, but with
$T_c\approx 0.3$, which is significantly lower than the
theoretical $T_c$.
\end{references}
\end{document}